\documentclass[10pt, conference, letterpaper]{IEEEtran}
\usepackage{graphicx}
\usepackage{amssymb, amsmath}
\usepackage{subfigure}
\usepackage[hyphens]{url}
\usepackage{epstopdf}
\usepackage{algorithm}
\usepackage{algpseudocode}
\usepackage{expl3}
\newcommand{\qualitiesnum}{m}
\newcommand{\mylabel}{y}

\newcommand{\naivefullname}{Nearest Neighbor using Average Bit Rate Feature}

\newcommand{\oursolfixed}[1]{Proposed solution, #1 training video titles (different streams), percentage of fixed qualities representation identification}

\newcommand{\testFixed}{\emph{test-fixed-train-titles}}
\newcommand{\testAdaptiveTrainTitles}{\emph{test-adaptive-train-titles}}
\newcommand{\testAdaptiveTestTitles}{\emph{test-adaptive-test-titles}}

\newcommand{\mycomment}[1]{}

\newcommand{\eg}{\emph{e.g.\@ }}

\newcommand{\etc}{etc.\@ }
\newcommand{\naive}{na\"{\i}ve}

\title{Real Time Video Quality Representation Classification of Encrypted HTTP Adaptive Video Streaming - the Case of Safari} 

\author{
\IEEEauthorblockN{Ran Dubin, Ofer Hadar, Itay Richman, Ofir Trabelsi} 
\IEEEauthorblockA{Communication Systems Engineering\\Ben-Gurion University of the Negev\\ Israel\\} 
\and
\IEEEauthorblockN{Amit Dvir} 
\IEEEauthorblockA{Center for Cyber Technologies\\Department of Computer Science\\ Ariel University\\ Israel\\} 
\and
\IEEEauthorblockN{Ofir Pele} 
\IEEEauthorblockA{Center for Cyber Technologies\\Department of Computer Science\\Department of Electrical and Electronics Engineering\\Ariel University\\ Israel\\} 

}

\begin{document} 
\maketitle 

\begin{abstract}
The increasing popularity of HTTP adaptive video streaming services
has dramatically increased bandwidth requirements on operator
networks, which attempt to shape their traffic through Deep Packet
Inspection (DPI). However, Google and certain content providers have
started to encrypt their video services. As a result, operators often
encounter difficulties in shaping their encrypted video traffic via
DPI. This highlights the need for new traffic classification methods
for encrypted HTTP adaptive video streaming to enable smart traffic
shaping. These new methods will have to effectively estimate the
quality representation layer and playout buffer. We present a new
method and show for the first time that video quality representation
classification for (YouTube) encrypted HTTP adaptive streaming is
possible. We analyze the performance of this classification method with
Safari over HTTPS. Based on a large number of offline and online
traffic classification experiments, we demonstrate that it can
independently classify, in real time, every video segment into one of
the quality representation layers with 97.18\% average accuracy.
\end{abstract}
\begin{IEEEkeywords}
HTTPS Video Streaming, Encrypted Traffic, Quality Representation Classification, Safari
\end{IEEEkeywords}

\section{Introduction}
\label{Introduction}
Every day, hundreds of millions of Internet users view videos online,
in particular on mobile phones whose numbers are clearly going to
increase\cite{Cisco_2, Cisco_zettabytes}. As a result, video streaming
is also expected to mushroom. For example, Google's streaming service,
YouTube, now occupies a market share of over $17\%$ of the total
mobile network bandwidth \cite{sandvine_2014, Cisco_zettabytes} in
North America. Google started a new user security revolution
by pushing the entire web traffic into HTTP Secure (HTTPS)
\cite{GoogleSSL} by giving a ranking boost in their search engine to
secure sites. As a result, YouTube network traffic is now encrypted.

Since online video streaming are fully viewed in less than $50\%$ of
the cases \cite{AdCreator} traffic shaping can reduce unnecessary
traffic waste. Network traffic classification algorithms use two main
techniques: DPI packet content analysis and statistical feature
classification
\cite{dainotti2012issues,valenti2013reviewing,cao2014survey,
  dubin2012progressive,niemczyk2014identification, ssl_clas, SSL_ext,
  korczynski2012classifying}. However, their effectiveness for
encrypted traffic is concentrated mainly in recognizing TLS/SSL
handshake parameters that help recognize the application content types
(video, chat, \etc) or the application name. They do not try to
classify the video stream quality representation or provide any
enrichment data on the video streams.


The YouTube video streaming solution is based on Adaptive Streaming
Over HTTP (DASH) \cite{DASH_RFC_1}. DASH is a Multi Bit Rate (MBR)
streaming method, designed to improve viewers' Quality of Experience
(QoE). In DASH, each video is divided into short segments, typically a
few seconds long ($2-16$ seconds), and each segment is encoded several
times, each time with a different quality representation level. The
user (player) adaptation logic algorithm is responsible for the
automatic selection of the most suitable quality representation for
each segment, based on the client's playout buffer and network
conditions. As a result, the quality representation layer in DASH can
change between segments. A content classification algorithm for
encrypted video streaming should recognize each quality representation
change. A video quality representation classification of encrypted
video streams can help in many ways such as collecting users' viewing
preferences, estimating the client playout buffer, tracking the users'
Quality of Experience (QoE) / Quality of Service (QoS). These are the
basic steps needed for designing video network traffic optimization
algorithms. These algorithms are used by the ISP for controlling its
network bandwidth.
	
In this paper we present a novel real-time video stream quality
representation classification for DASH. We classify the video quality
representation, and each feature (group of packets) is classified by
itself without any dependencies on past or future samples. Our scheme
was tested on the Safari browser with Adobe flash as the player over
HTTPS network traffic on offline and online YouTube video traffic
streams. It recognizes, in real time, the YouTube video traffic quality
representation layer with $97.18\%$ average accuracy. Our method can
also be used for estimating the client's playout buffer and as a basic
step in traffic shaping.
 
The remainder of this paper is organized as follows. In Section
\ref{Related Works} we discuss related work. YouTube analysis is
presented in Section \ref{YouTube_Analysis}. The problem formulation
is introduced in Section \ref{Problem Formulation}. Section
\ref{Proposed Algorithm} presents our new algorithm. Section
\ref{Performance Evaluation} presents the performance
evaluation. Finally, section \ref{Conclusions} discuss our conclusions
and future work.

\begin{figure*}[htbp]
        \centering
		\subfigure[Firefox auto mode over HTTP$2$.]{\label{fig:videoCapture:ff_auto}\includegraphics[width=0.45\textwidth]{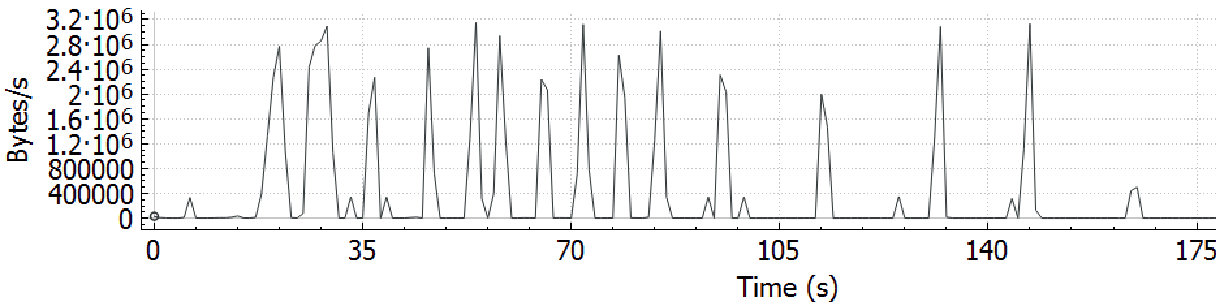}}
		\subfigure[Firefox fixed mode over HTTP$2$]{\label{fig:videoCapture:FF_fixed}\includegraphics[width=0.45\textwidth]{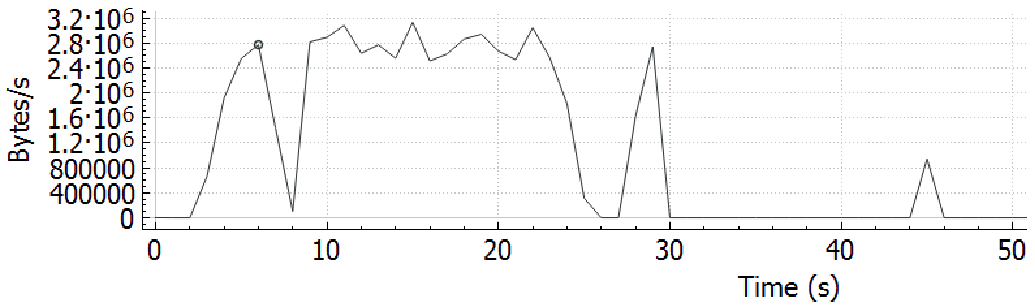}}
		\subfigure[Safari auto mode over HTTPS]{\label{fig:videoCapture:safari_auto}\includegraphics[width=0.45\textwidth]{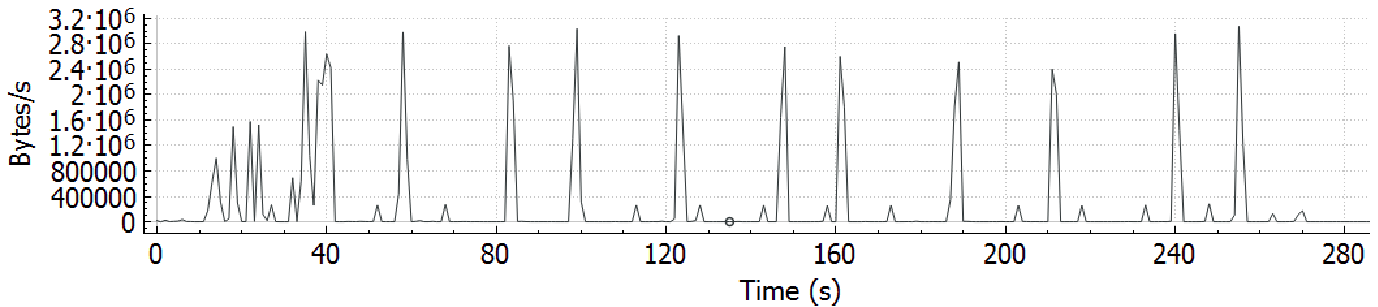}}
		\subfigure[Safari fixed mode over HTTPS]{\label{fig:videoCapture:safari_fixed}\includegraphics[width=0.45\textwidth]{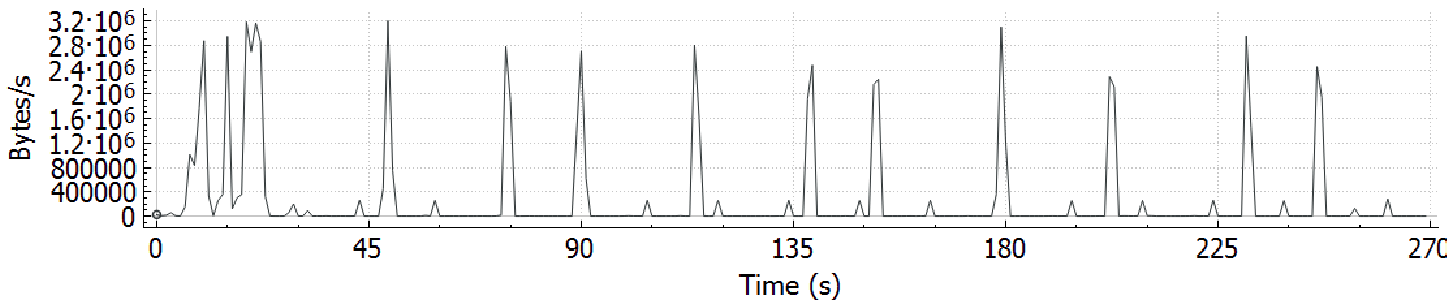}}
		\subfigure[Explorer auto mode over HTTPS]{\label{fig:videoCapture:explorer_auto}\includegraphics[width=0.45\textwidth]{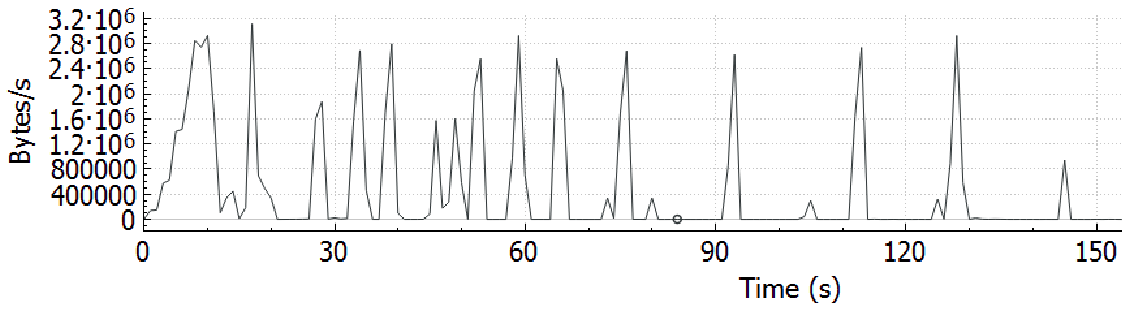}}
		\subfigure[Explorer fixed mode over HTTPS]{\label{fig:videoCapture:explorer_fixed}\includegraphics[width=0.45\textwidth]{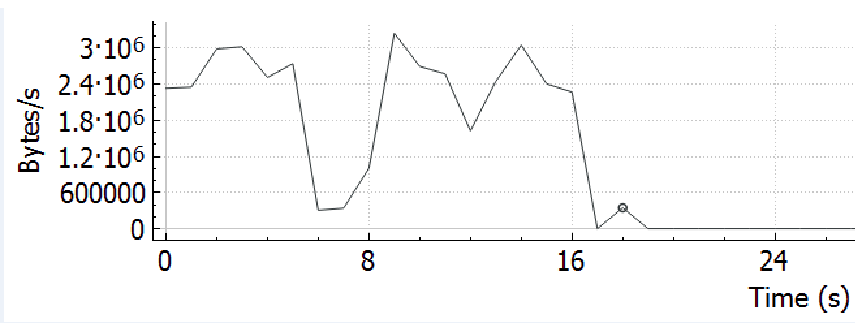}}
		\subfigure[Chrome auto mode over HTTP$2$]{\label{fig:videoCapture:chrome_auto}\includegraphics[width=0.45\textwidth]{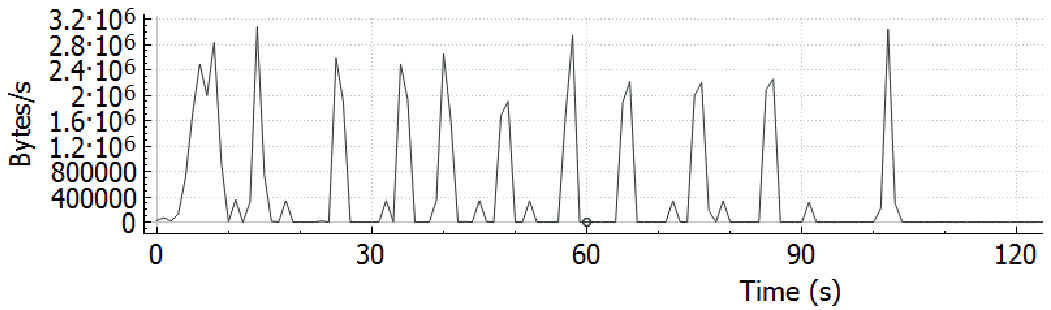}}
				\subfigure[Chrome fixed mode over HTTP$2$]{\label{fig:videoCapture:chrome_fixed}\includegraphics[width=0.45\textwidth]{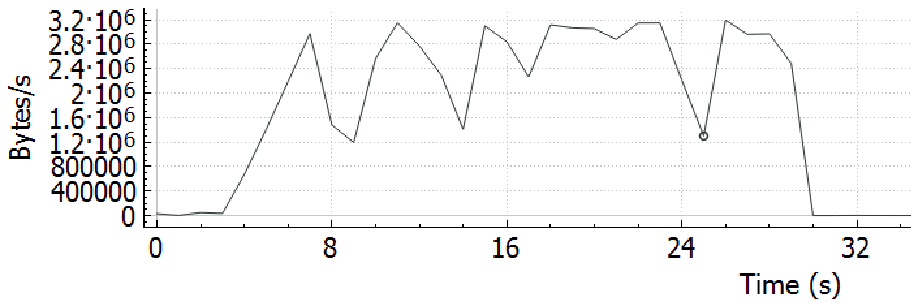}}
		    \caption{YouTube Costa Rica in 4K - traffic traces from different
              browsers: Safari (Windows Ver $5.1.7$) with flash player
              , Firefox (Ver $37$) with $HTML5$ player, Explorer (Ver
              $11.0.96$) with $HTML5$ player and Chrome (Ver
              $43.0.2357.81$) with $HTML5$ player.}
        \label{fig:videoCapture}
\end{figure*}

\begin{figure*}[htbp]
   \centering
		\subfigure[Video and audio flows]
	{\label{fig:safari_auto_video_audio}
	\includegraphics[width= 15cm, height=2cm]{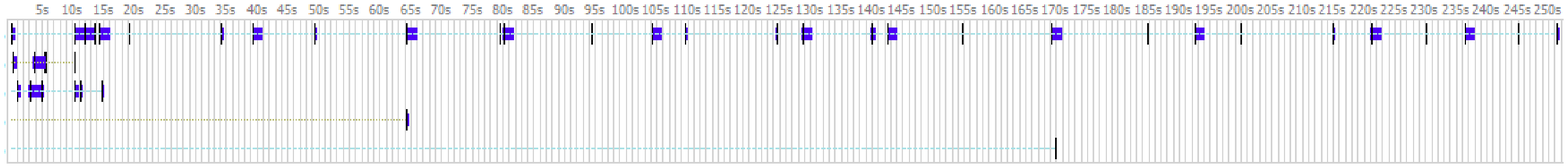}}
	
	\subfigure[Video flows (without audio)]
	{\label{fig:safari_auto_video_only}
	\includegraphics[width= 15cm, height=2cm]{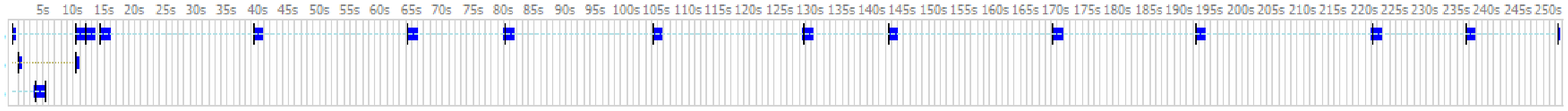}}

\caption{YouTube Costa Rica 4k auto mode with Safari. Each horizontal line represents
  different YouTube flows from the same download. The video quality changes from $360P$ to $720P$.}
\label{fig:costaRica}
\end{figure*}

\section{Related Work}
\label{Related Works}
Many recent works have suggested methods for encrypted traffic
classification and several surveys have presented detailed description
of the state of the art methods
\cite{dainotti2012issues,valenti2013reviewing,cao2014survey}. Several
works have examined different statistical features such as session
duration \cite{paxson1994empirically,alshammari2010unveiling,
  zander2005self}, number of packets in a session
\cite{alshammari2010unveiling, zhang2010identification,
  paredes2012practical}, different variance calculations of the
minimum, maximum and average values of inter-arrival packet time
\cite{alshammari2010unveiling,zhang2010identification}, payload size
information \cite{zhang2010identification,bonfiglio2009detailed}, bit
rate \cite{bonfiglio2009detailed, chen2006quantifying}, Round-Trip
Time (RTT) \cite{chen2006quantifying}, packet direction
\cite{hjelmvik2009statistical} or server sent bit rate
\cite{bar2010realtime}. Not all these features are important for video
streams classification. For instance, the packet size is often MTU size in
video streaming, as video streaming consumes high bandwidth and
re-transmission occurs often. Moreover, TCP parameters such as server
sent bit rate, inter-arrival packet time, RTT and packet direction are
weak features.  Other classification methods have identified the
application type and class (VOIP, Video, \etc)
\cite{dainotti2012issues,valenti2013reviewing,cao2014survey},
by exploiting encrypted VOIP streams interaction of Variable Bit Rate
(VBR) codecs such as phonetic reconstruction
\cite{white2011phonotactic} and language identification
\cite{wright2007language}. However, these methods need many trace
samples for the training of their classification models.

Malware traffic fingerprinting methods were suggested by Siboni et
al. \cite{SiboniCohen2014} and Shimoni et
al. \cite{shimonimalicious}. Both methods are based on the Lempel Ziv
$78$ ($LZ78$) universal compression algorithm \cite{Lempel1978} and on
probability tree classifiers. First, a statistical feature based on
time differences of all the training samples is created, quantized and
transformed into a discrete sequence over small finite alphabet (a
single code-book for all trees). In the next step, the sequence is
used for building a $LZ87$ tree for each training sample with a
probabilistic prediction model \cite{SiboniCohen2014}. In the testing
phase, a similar process is activated and tested with the training
database trees. Malware fingerprinting is not designed for use in our
case. Therefore, we modified the Shimoni et al.  algorithm
\cite{shimonimalicious} to the streaming world. We used this modified
algorithm as one of the methods against which our method is compared.

In this work, we use the client's received bit rate with TCP stack
implementation to overcome re-transmissions. We show that using
time-based features for video streaming leads to poor classification
results. The objective of this work is to develop a real time classifier
for the encrypted video traffic quality representation layer and web
browsers, solutions that cannot classify every segment's quality
representation by itself are not suitable. Rather, our proposed solution is a
stream based classification method.
\section{YouTube Analysis}
\label{YouTube_Analysis}
To better understand encrypted video streaming traffic properties, we
examined YouTube traffic under different browsers.  In
Fig. \ref{fig:videoCapture} depicts the different traffic download
patterns of a single video stream. In each download, we used the same
video stream with a fixed quality representation of $720P$ over
different browsers. The different traffic patterns are mainly caused
by the browsers' player algorithms. However, the source video encoding
process also affects pattern differences. It is noteworthy that at the
time of our database creation, Explorer and Chrome had YouTube
$HTML5$ players while Firefox and Safari had a Flash based player.

Fig. \ref{fig:videoCapture} shows that $HTML 5$
players in the fixed quality representation mode and Adobe flash players have
significantly different traffic patterns. The flash traces and $HTML5$
players in the automatic mode have high bursty traffic with
a silence separation of around $3$ seconds between peaks, whereas the
$HTML5$ traffic has one high and short traffic
burst. Chrome downloaded a video stream
with a duration of $281$ seconds in less than $30$ seconds. As a
result, different feature extraction methods are needed to
identify the different players' requested streams.

\begin{figure}[htbp]
\centering{\includegraphics[width= 8cm, height=8cm]{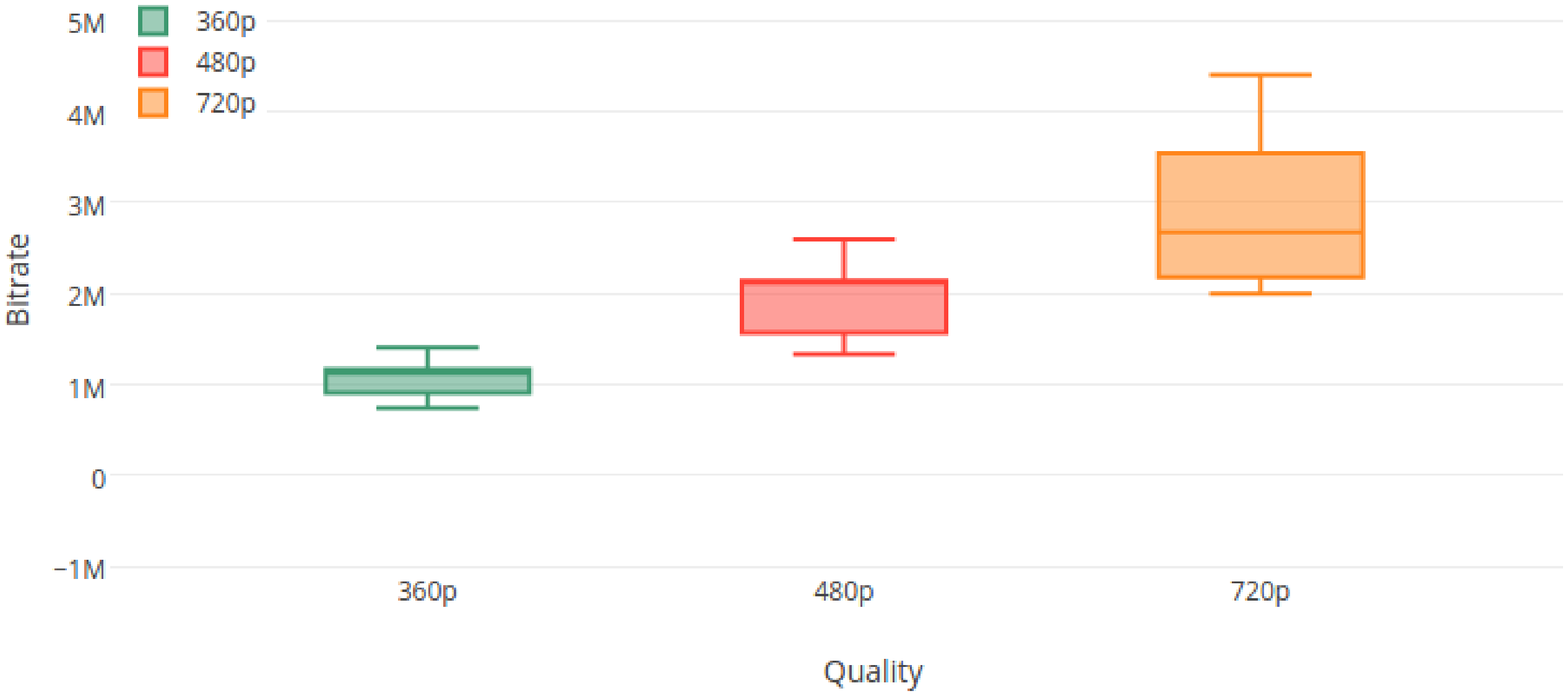}}
\caption{Safari dataset network bit rate Vs quality representation confidence ($95\%$ level)}
\label{fig:safari_confidence}
\end{figure}

Fig. \ref{fig:safari_auto_video_audio} illustrates the YouTube
automatic download mode with Safari. Each video download has several
flows. In the Safari fixed quality representation, there is one main
video flow and 3-5 parallel flows (including audio only flows). Some
of the flows can be used for downloading the same quality
representation in parallel to accelerate the download. By using
the Fiddler \cite{lawrencefiddler} web debugging proxy we can view the
different requests without the encryption. The small
traffic peak periods are the audio while the video peaks take longer
to download. This analysis leads to several insights concerning the
factors that can hinder classification efforts:
	
\begin{enumerate}

\item The audio data and the video data can be found in the same
  5-tuple flow and in some cases we cannot distinguish between
  them. This can result in a classification error since the
  boundaries between the quality representations are very close (see
  Fig. \ref{fig:safari_confidence}), which illustrates the dataset
  confidence graph for each of the tested quality representations. In
  Fig. \ref{fig:safari_confidence}, the $360P$, $480P$ and
  $720P$ have overlapping bandwidth ranges. This makes the
  classification effort harder. This can also be seen in the first
  flow (Fig. \ref{fig:safari_auto_video_audio}) at $14$ seconds where
  the audio download is very close to the video traffic before and
  after. As a result, we cannot distinguish between them in their
  network traffic representation.

\item Close video segments' responses can be found in the same flow. For example, in
  Fig. \ref{fig:safari_auto_video_audio} in the first flow ($11-14$
  seconds) there are two downloaded segments that have very small time
  differences between the responses in the encrypted traffic
  representation. Distinguishing between segments that were downloaded
  at $11-14$ seconds is difficult.

\item The first segment in each flow has a high bit rate variance
  which in most cases is not unique to a specific quality
  representation. For this reason we chose not to use it in training
  and testing.

\item The last segment usually consist of data leftovers. Its behavior
  is different, hard to predict and its classification is less
  important since this is the end of the stream. Hence this segment
  was not used.
\end{enumerate}

After filtering the audio responses
(Fig. \ref{fig:safari_auto_video_only}) it can be seen that up to the
first $10$ seconds the $360P$ quality representation was downloaded in
parallel. Afterward, there was a new parallel download for the $720P$
quality representation. These qualities were observed in the Fiddler
traces but other traces evidenced additional quality representation
switching.

\begin{figure*}  
\subfigure[QUIC $360P$ fixed quality representation download]{\label{fig:QUIC_Capture_fixed_360}\includegraphics[width=0.33\textwidth]{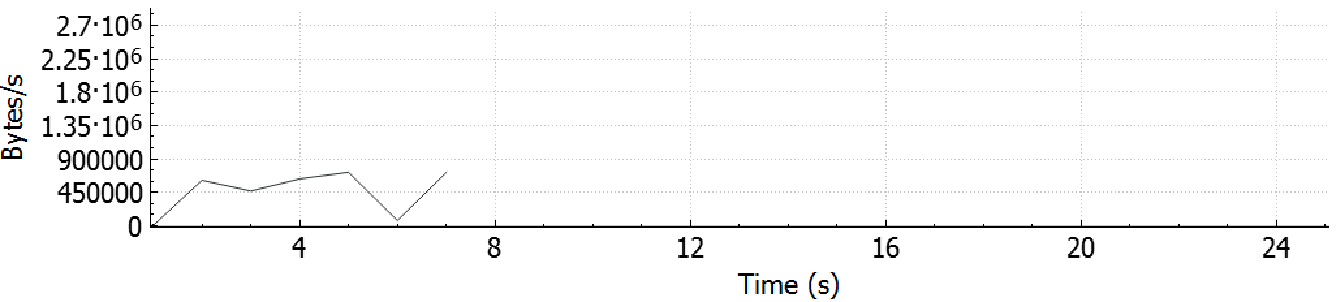}}
\subfigure[QUIC $480P$ fixed quality representation download]{\label{fig:QUIC_Capture_fixed_480}\includegraphics[width=0.33\textwidth]{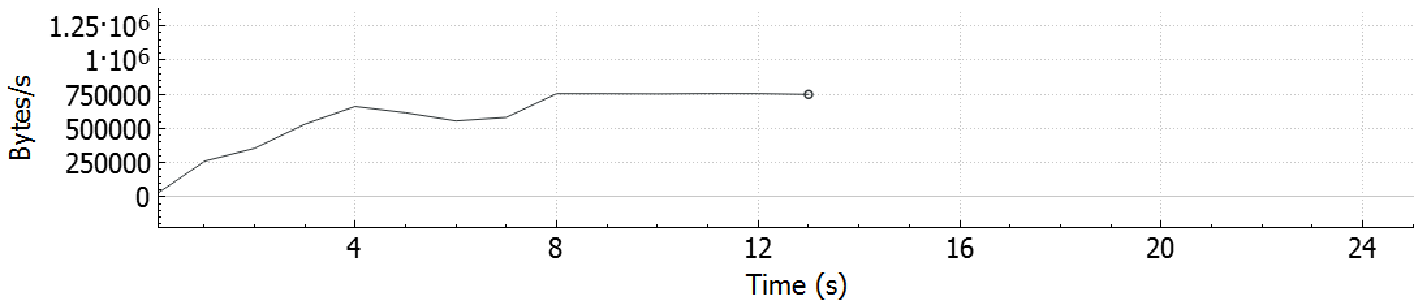}}
\subfigure[QUIC $720P$ fixed quality representation download]{\label{fig:QUIC_Capture_fixed_720}\includegraphics[width=0.33\textwidth]{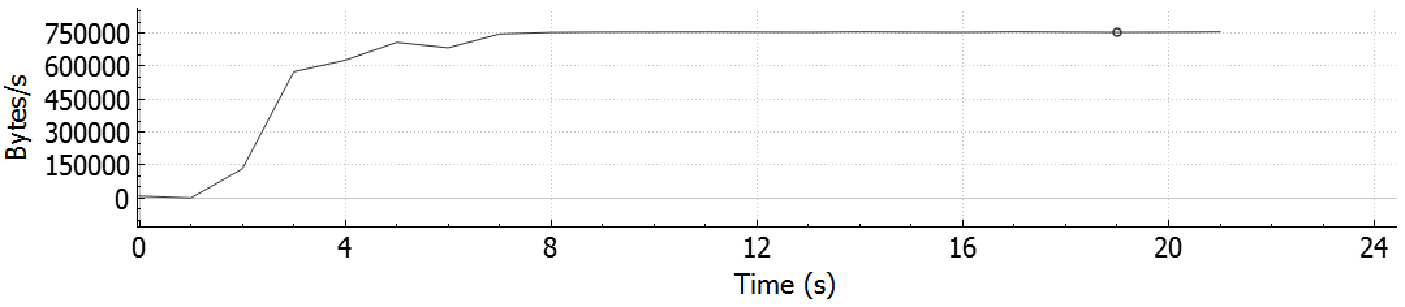}}
\subfigure[QUIC auto mode]{\label{fig:QUIC_Capture_fixed_auto_mode}\includegraphics[width=0.45\textwidth]{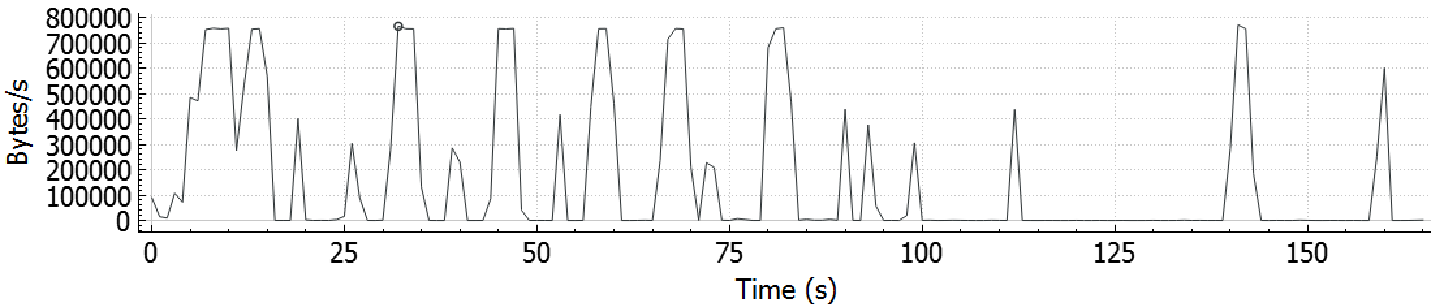}}
\caption{QUIC over UDP - traffic traces from a Chrome browser with different fixed and auto quality representations}
  \label{fig:QUIC_Capture}
\end{figure*}

YouTube in Chrome can be downloaded not only with HTTP$2$/SPDY and
HTTPS but also with QUIC over UDP. Fig. \ref
{fig:QUIC_Capture_fixed_360}-Fig. \ref{fig:QUIC_Capture_fixed_720}
illustrates the download of the same video with the following quality
representations: $\{360P, 480P, 720P\}$ with QUIC. The download
throughput in this case is similar but the download duration is longer because quality
representation is higher. The QUIC auto mode behavior, plotted in
Fig. \ref{fig:QUIC_Capture_fixed_auto_mode} is similar to
HTTP$2$ behavior.

We decided to focus on Safari, due to the fact that the fixed and auto
mode have similar behavior. QUIC throughput characterization would be 
interesting for future work. After many experiments, we found that between
the end of one traffic burst and the next there is a time window exceeding
$3$ seconds of silence. Thus henceforth we
define bit rate as bit per peak (traffic burst).

\section{Problem Formulation}
\label{Problem Formulation}

A server stores a video which is segmented into fixed duration
segments. Each segment is encoded into $\qualitiesnum$ representations
($\qualitiesnum$ can be different for different videos). The user can
select to download a constant or adaptive representation download. In
the adaptive mode, the client's video player application (via adaptation
logic), based on his network condition estimate and playout buffer
selects a suitable representation to download each segment.

We used data from static (constant) quality representations to learn a
model that can classify segments of constant and adaptive video
streams.  We used a training set of encrypted video streams, where
each was downloaded $\qualitiesnum$ times. Each download had different
constant video quality representation. We used a fixed
$\qualitiesnum=3$ for all videos. Every segment of the stream is
encoded to a feature. The label of each segment is its constant
quality representation index: $\mylabel \in \{1 \ldots m\}$ (\eg 1 for
$360P$, 2 for $480P$ and 3 for $720P$). In the next section we
describe our encoding of a stream segment into a feature vector and
how we learn a model that can classify stream segments.

\section{Proposed Algorithm}
\label{Proposed Algorithm}

\begin{figure}[htbp]
\centering{\includegraphics[width= 8cm, height=6cm]{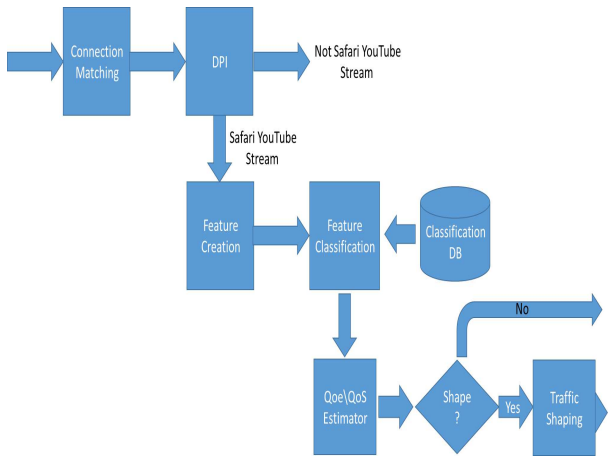}}
\caption{Proposed solution architecture}
\label{fig:generalSolutionDiag}
\end{figure}

The proposed solution architecture is illustrated in
Fig. \ref{fig:generalSolutionDiag}. The first two modules only pass
YouTube video streams to the next modules. Each segment of network
traffic enters the system separately and is first passed into the
\textit{Connection Matching} filter. This filter is responsible for
checking whether the incoming flow is new or ongoing. It does so based
on a five-tuple representation: \{protocol (TCP/UDP), src IP, dst IP,
src port, dst port\}. If the incoming flow is new, the \textit{DPI}
filter decides whether it is a YouTube flow. This is done based on the
Service Name Indication (SNI) field in the \emph{Client Hello}
message. If the \textit{DPI} module finds the following
string: \textit{googlevideos.com} (which identifies YouTube) in the SNI, the
stream is passed to the \textit{Feature Creation} module. Any ongoing
or new traffic flow that is not recognized by the \textit{DPI} as
video streaming is transparently passed into the network without
further analysis. Note that in this paper we assume that
we know how to detect Safari browser traffic (in contrast to other
browser traffic). This can be done by identifying the audio stream of
Safari. This task is left for future work.

The \textit{Feature Creation} module extracts statistical features in
real time based on the arriving packets (see section \ref{Feature
  Creation}). The \textit{Feature Classification} module classifies
the quality representation (see section \ref{Traffic Feature
  Classification Engine}).

Finally, the \textit{QoE/QoS Estimator} module predicts the client
playout buffer and estimates re-buffering events. This information is
needed for the shaping of the encrypted traffic. The \textit{QoE/QoS Estimator} and shaping modules are
left for future work.
\subsection{Feature Creation}
\label{Feature Creation}
DASH is streamed over a TCP transport protocol. Streaming applications
have high bit rate consumption. Thus, feature creation methods need to
take TCP limitations such as re-transmission caused by
network problems into account.  Re-transmission adds additional data to the stream
that can cause classification errors.

In section \ref{Related Works}, we discussed state-of-the-art network
traffic feature creation methods such as packet length, inter-arrival
packet time and RTT packet direction. However, as the payload size in
video streaming is often maximum size, delays in the network are varied
and re-transmissions cause false packet counts. Therefore, we suggest a
single dimension bit rate feature based on a TCP stack re-transmission
filter using the TCP ACK method.

The feature creation starts after we identify that this traffic flow
is a YouTube video flow. Any packet that enters the algorithm is
verified by TCP stack implementation to prevent re-transmission
packets from affecting our feature accuracy. We found $3$ to be a good
traffic feature threshold. We ignored low bit rate traffic features
that can represent audio traffic bursts.

\subsection{Feature Classification}
\label{Traffic Feature Classification Engine}

The proposed classification solution is illustrated in
Fig. \ref{fig:algorithmDiagram}. It has a training step and a testing
step. In the training step, first, we constructed our dataset based on
YouTube video streaming captures (PCAP trace files \cite{ourDB}). Each video was
downloaded with the three following fixed qualities $\{360P, 480P,
720P\}$. In the second stage, we extracted statistical features from the
entire labeled data-set. In our proposed solution the statistical
feature is a bit rate throughput in a time period based on the user's TCP stack
implementation which filters out unnecessary TCP re-transmissions that
occurred regularly in the traffic. Our feature extraction method
is customized to the browser generated content (Safari). In the third stage,
the entire features set was clustered using $k$-means++
\cite{arthur2007k} (step (3) in Fig. \ref{fig:algorithmDiagram}). The end product
of these steps is a single dimension code-book that represent the
entire feature set. 

For each quality, we iterated over all its traces and averaged every
peak total bit rate. This yielded an average bit rate vector for
each quality.  From these vectors and using the codebook from the
$k$-means stage we computed a representative string for each quality.
In the classification stage we carried out the bit rate extraction for each
segment and then assigned a symbol (the one with the shortest distance
to the average) to it from the codebook. Finally we assigned a label
by finding which center was the closest.

\section{Performance Evaluation}
\label{Performance Evaluation}

\begin{figure*}[htbp]
\centering{\includegraphics[width= 14cm, height=8cm]{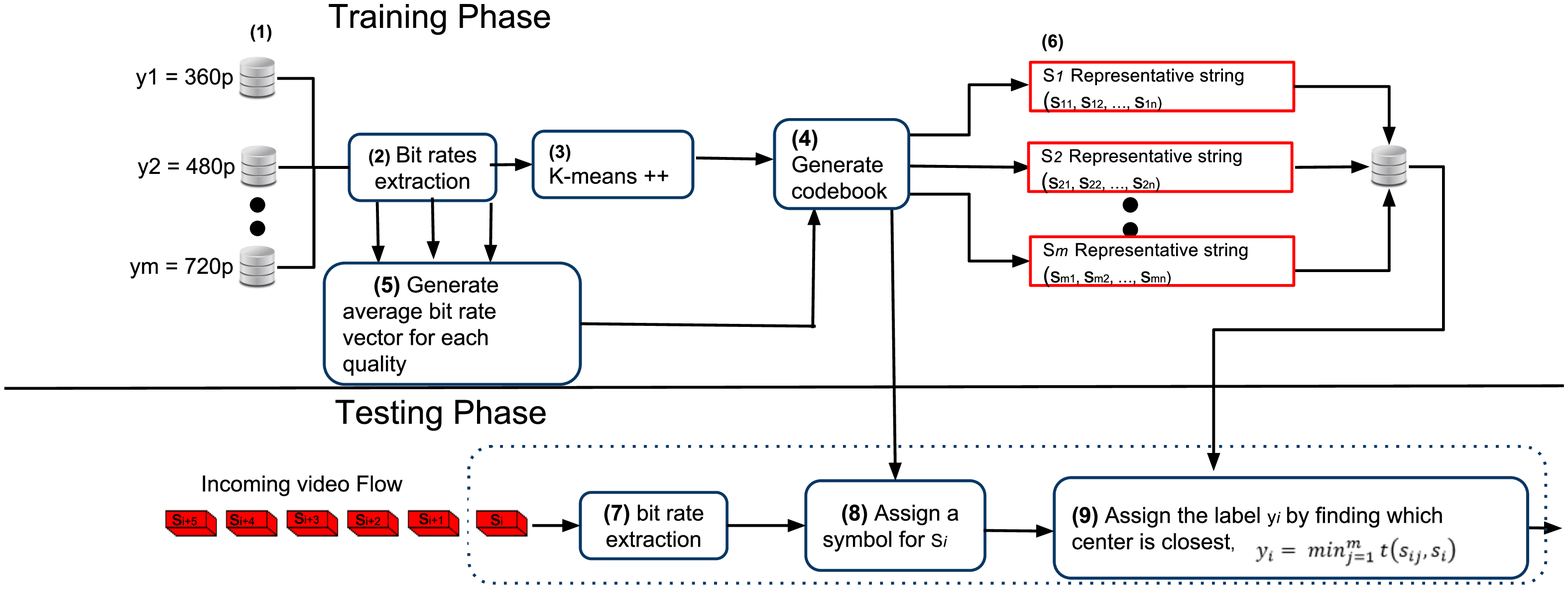}}
\caption{Proposed algorithm diagram flow}
\label{fig:algorithmDiagram}
\end{figure*}

In this section, we evaluate the proposed quality representation classification algorithm. First, we describe the dataset in \ref{Datasets}. Then we analyze the accuracy with different numbers of
$k$-means centers (step (3) in Fig. \ref{fig:algorithmDiagram}) in
Section \ref{Evaluation of the optimal $k$ and data size
  discussion}. In Section \ref{Evaluation of accuracy using different
  training dataset sizes} we evaluate the accuracy using different
training dataset sizes. We analyze the accuracy on the different test
sets in Section \ref{Accuracy evaluation of auto/fixed representation
  mode selection}. We test the classifier's robustness to delays and
packet losses in Section \ref{Evaluation of robustness to delays and
  packet losses}. We examine the user buffer estimate accuracy in
Section \ref{User buffer estimation}. Finally, we compare our
classification results to two different classifiers in Section
\ref{Classifiers comparison}, one of which is a \naive{} algorithm we developed
and the other based on a malware anomaly detection algorithm
\cite{SiboniCohen2014,shimonimalicious} which we modified to the
streaming world.

\subsection{Dataset}
\label{Datasets}

The video titles used in this study are popular YouTube videos from
different categories such as news, video action trailers and GoPro
videos \cite{ourDB}. 

In this study we decided to focus on the Safari browser since the fixed
quality download mode (Fig. \ref{fig:videoCapture:safari_fixed}) and
the adaptive quality selection mode
(Fig. \ref{fig:videoCapture:safari_auto}) have similar
characteristics. We show that for Safari, we can learn an accurate
model for static or automatic quality modes simply by using a fixed
training dataset. Future studies will add additional browsers.

The training dataset contained $120$ video streams of $40$ unique
video titles each of which was separately downloaded with fixed quality from the
following qualities: $\{360P, 480P, 720P\}$.

We have three testing datasets:
\begin{enumerate}
\item \testFixed: $120$ video streams of $40$ unique video titles
  (same titles as in the training phase) each of which was separately downloaded
  with a fixed quality from the following qualities: $\{360P, 480P,
  720P\}$.
\item \testAdaptiveTrainTitles: $5$ video streams of $5$ unique video
  titles (titles taken from the training phase titles) each of which was
  downloaded with an adaptive quality representation (auto mode).
\item \testAdaptiveTestTitles: $5$ video streams of $5$ unique video
  titles (new titles that were not in the training phase) each of which was
  downloaded with an adaptive quality representation (auto mode).
\end{enumerate}
All the test video streams were different from the ones that were used
in the training phase (because of network conditions).


\subsection{Accuracy Evaluation using Different Numbers of $k$-means Centers}
\label{Evaluation of the optimal $k$ and data size discussion}

\begin{figure}[htbp]
\centering{\includegraphics[width= 8cm, height=5cm]{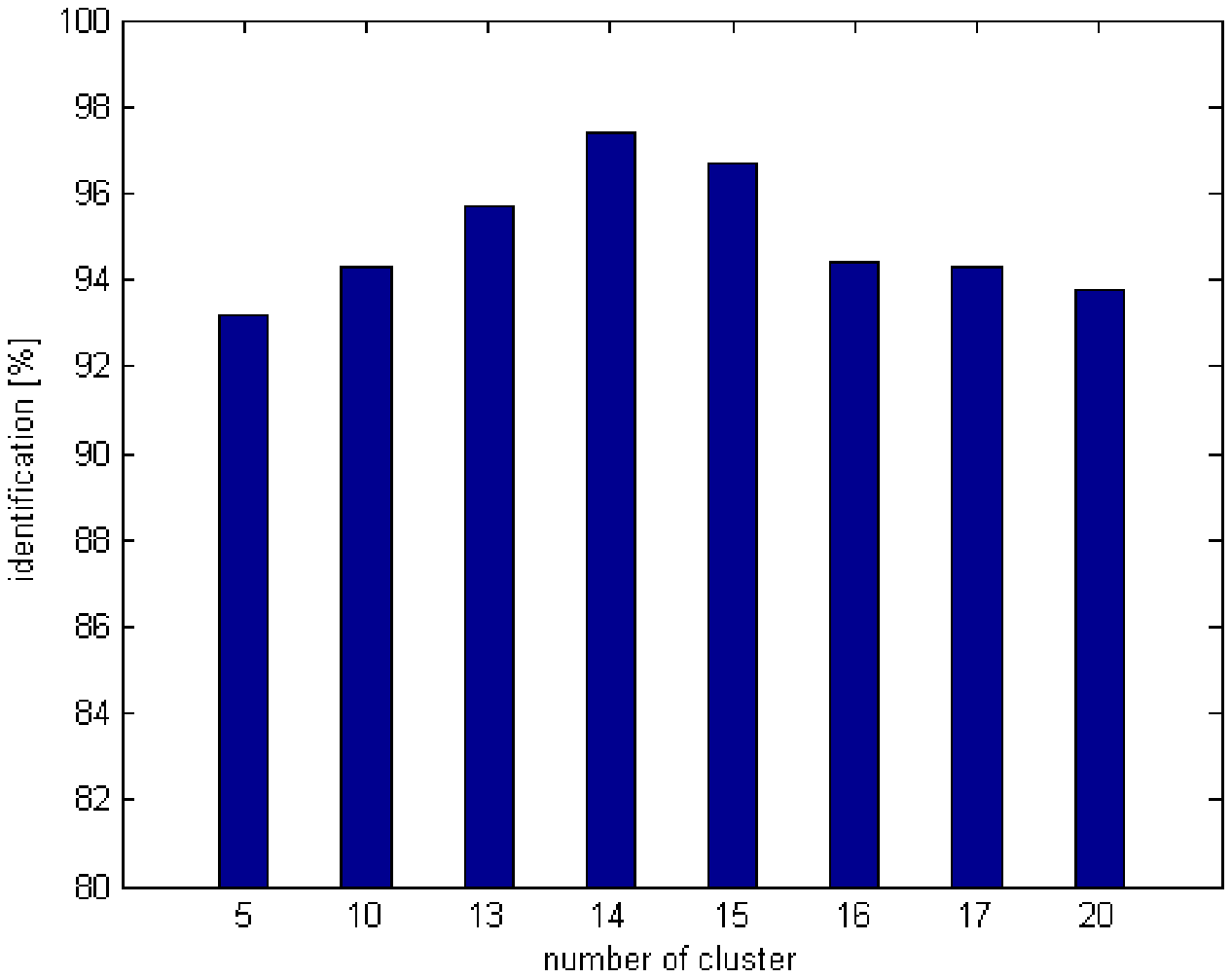}}
\caption{\oursolfixed{40} using different
  $k$ values (number of $k$-means centers)}
\label{fig:safari_clusterSize}
\end{figure}

Our solution cluster the bit rates into $k$ bins.  We tested the
classifier with our training dataset (see
Fig. \ref{fig:safari_clusterSize}). We found that $k = 14$ achieved
the highest classification accuracy and this is the $k$ that we used
in all the following experiments.

\begin{figure}[htbp]
\centering{\includegraphics[width= 8cm, height=6cm]{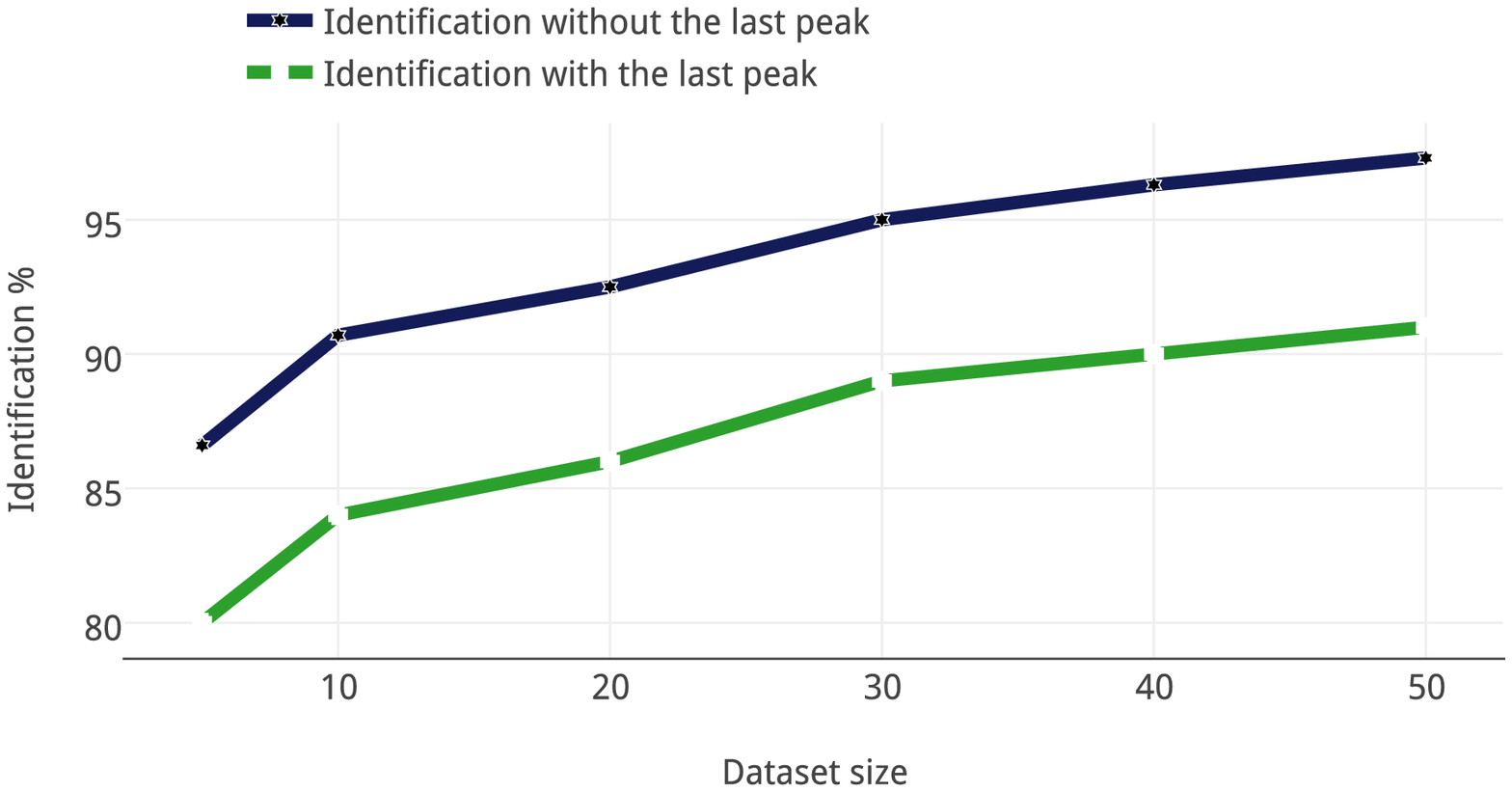}}
\caption{\oursolfixed{5-50}.}
\label{fig:safari_TP_vs_db_size}
\end{figure}


\subsection{Accuracy Evaluation using Different Training Dataset Sizes}
\label{Evaluation of accuracy using different training dataset sizes}

In Fig. \ref{fig:safari_TP_vs_db_size} we compare our recognition
identification rate with different numbers of training video titles.
The figure shows major gains in performance when the number of training video titles
 increases from $10$ to $30$. The gains are much smaller when
training video titles number increases from $30$ to $50$ (by only
$2.2$\%). The figure also shows that using the last peak in our
solution decreases the identification rate. The last peak size varies
(because it corresponds to the stream leftovers) and thus it decreases the
identification rate.


\begin{figure*}[htbp]  
  \subfigure[Proposed solution, 40 training video titles (different streams), fixed qualities representation.]{\label{fig:safari_confusion_matrix_our_fixed}
    \includegraphics[width=0.23\textwidth]{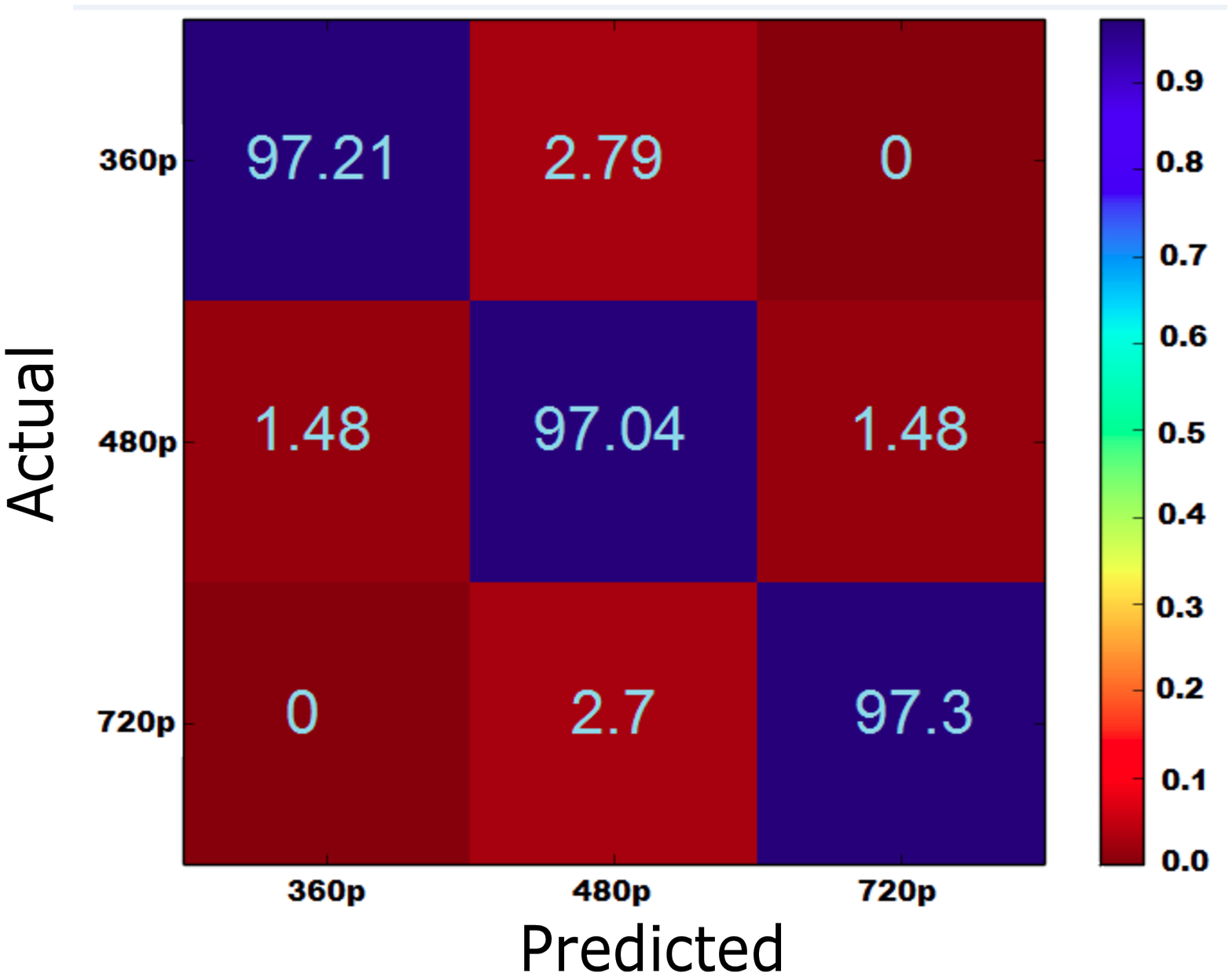}}
  \subfigure[\naivefullname (\naive), \testFixed: 40 training video titles (different streams), fixed qualities representation.]{
    \label{fig:safari_conf_without_algo}
    \includegraphics[width=0.23\textwidth]{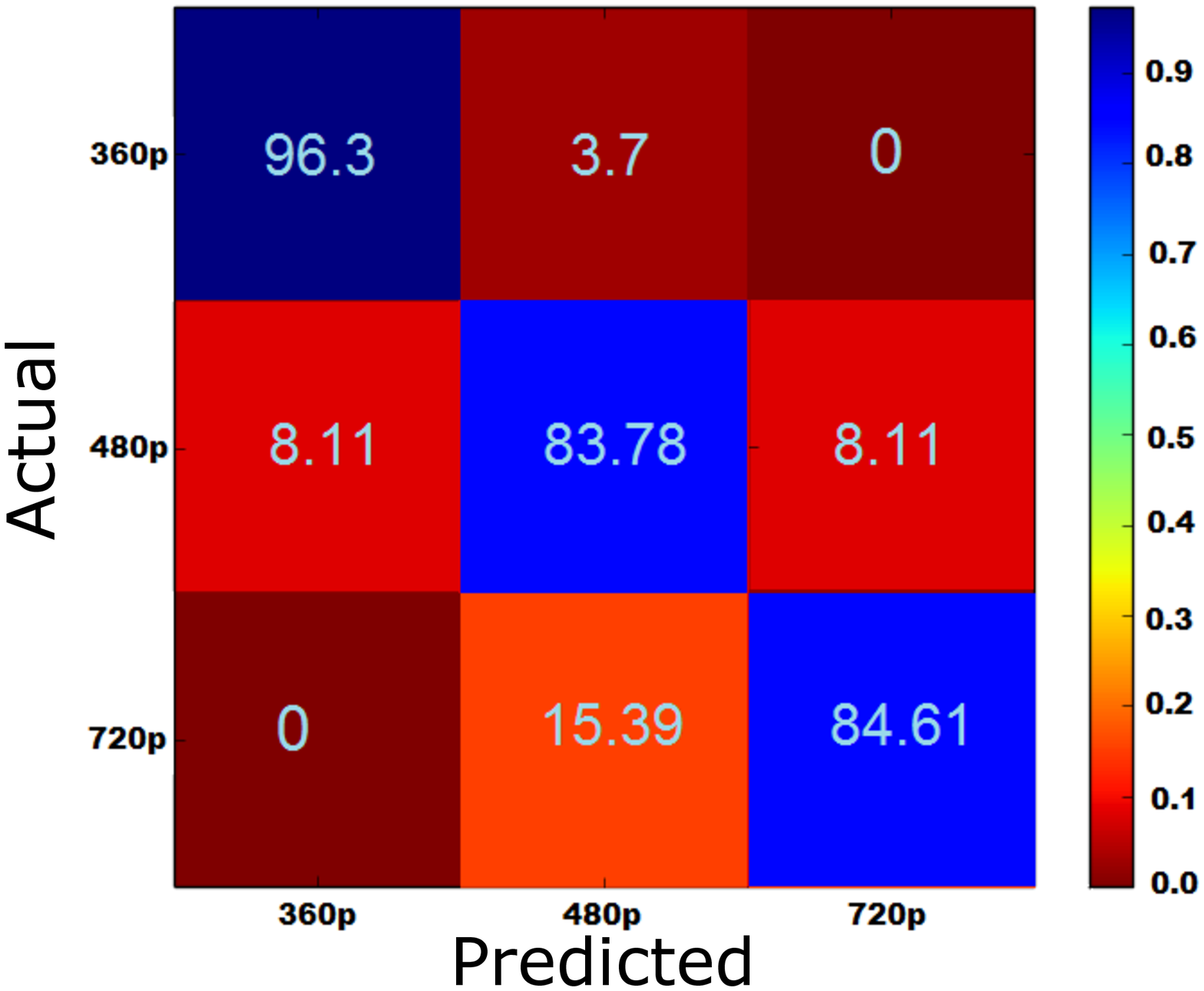}}
  \subfigure[Proposed solution, \testAdaptiveTrainTitles: 5 training video titles (different streams), auto quality representation.]{
	\label{fig:safari_conf_auto_from the data set}
    \includegraphics[width=0.24\textwidth]{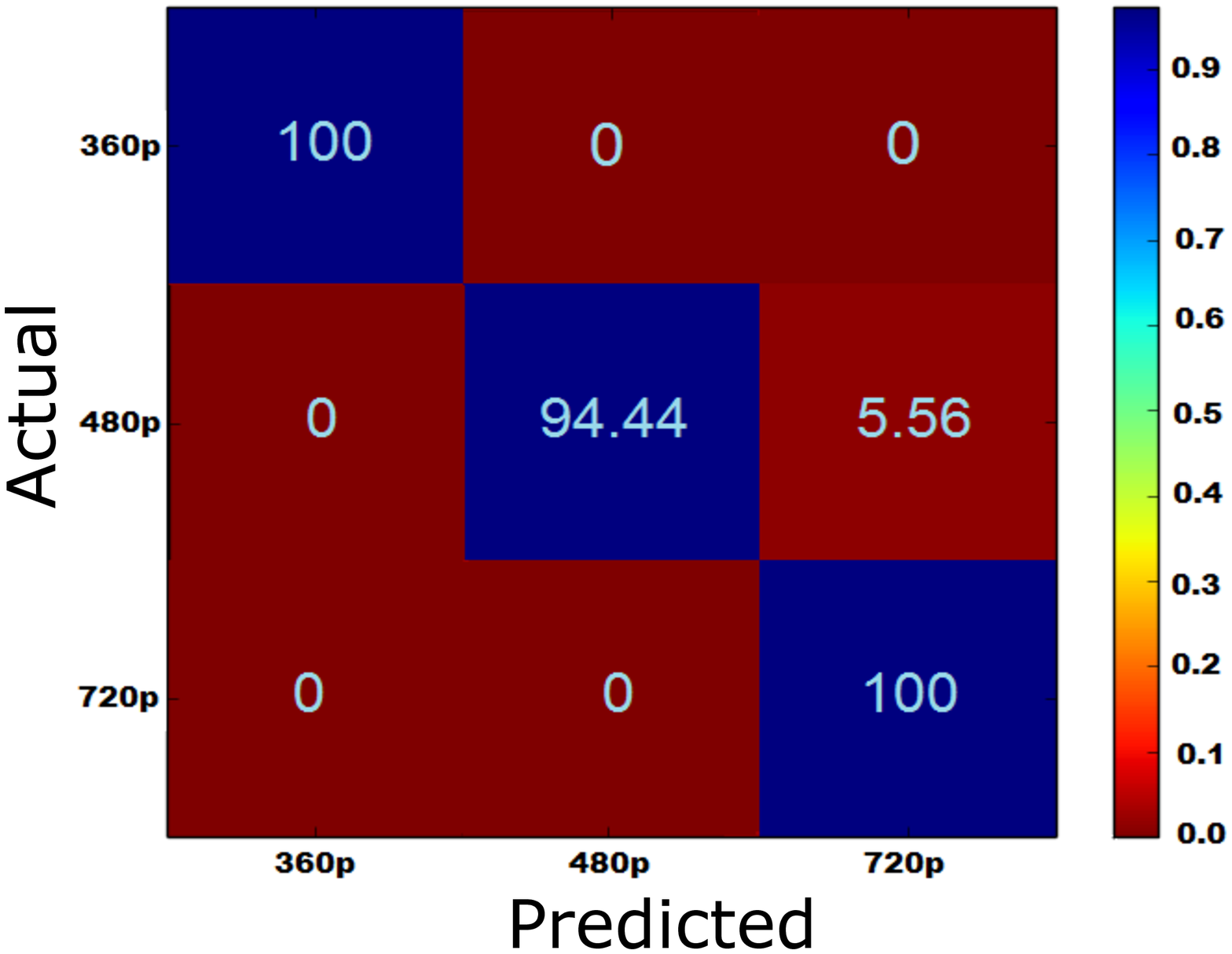}}
  \subfigure[Proposed solution, \testAdaptiveTestTitles: 5 new videos titles (not seen in training), auto quality representation.]{
    \label{fig:safari_confusion_matrix_auto_not_from_dataset}
    \includegraphics[width=0.23\textwidth]{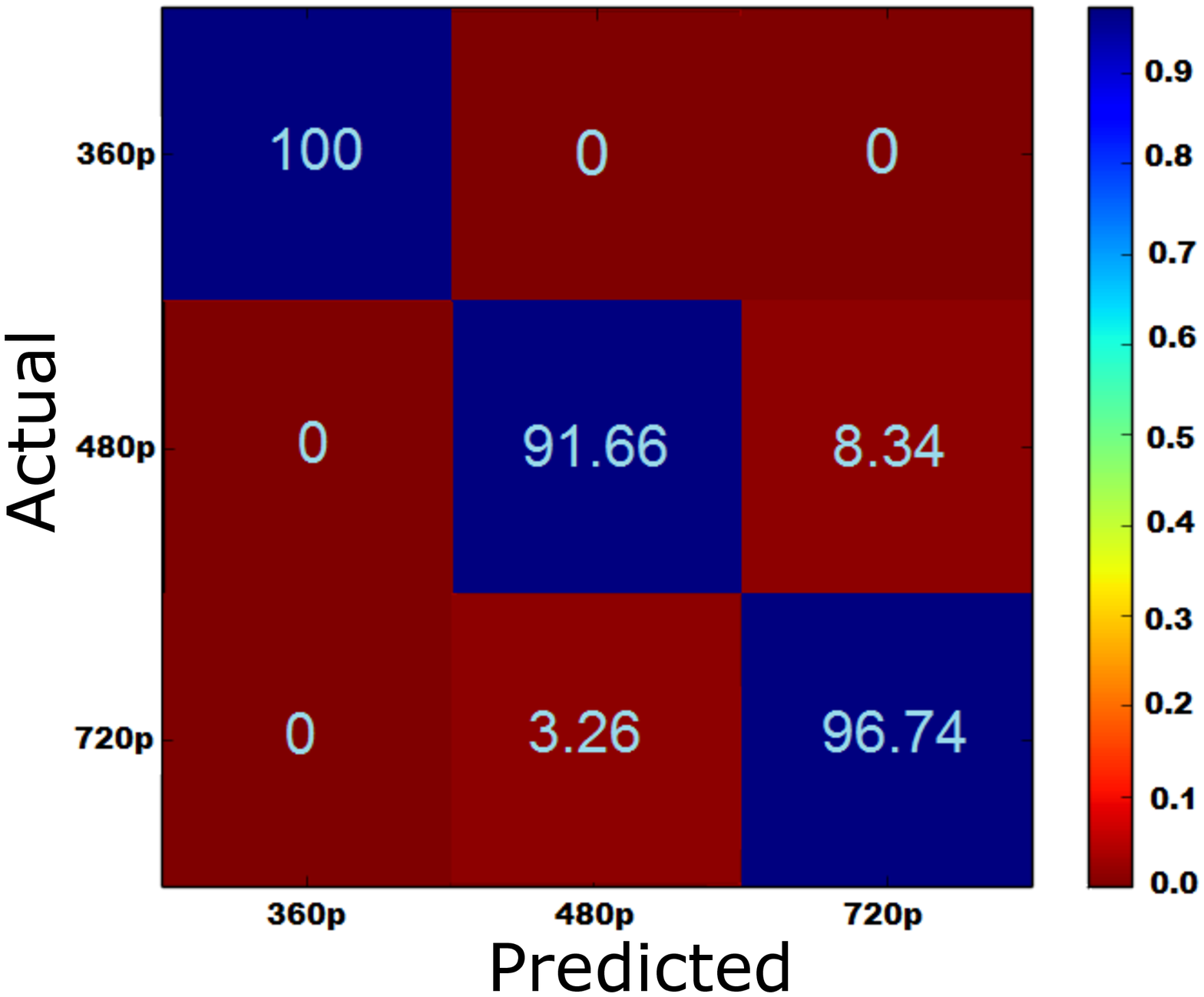}}
  \label{Safari_confusion_matrixes}
  \caption{Confusion matrices.}
\end{figure*}

\subsection{Accuracy Evaluation on the Different Test Sets}
\label{Accuracy evaluation of auto/fixed representation mode selection}

Fig. \ref{fig:safari_confusion_matrix_our_fixed} shows that our
classification errors in the fixed quality representation mode, are between
close quality representations and were lower than $3$\%. Note
that Fig. \ref{fig:safari_conf_without_algo} is based on the
\naivefullname (\naive) which uses the average bit rate that was
calculated from Fig. \ref{fig:safari_confidence} for each quality.

The average classification accuracy was $2 \%$ better when we tested
video titles from our training set
(Fig. \ref{fig:safari_conf_auto_from the data set}) than when we
tested video titles that were not in our training set
(Fig. \ref{fig:safari_confusion_matrix_auto_not_from_dataset}).

We examined why the error of classifying $480P$ quality representation
segments as $720P$ in adaptive streams was relatively higher than the other
errors (see Figs. \ref{fig:safari_conf_auto_from the data set} and
\ref{fig:safari_confusion_matrix_auto_not_from_dataset}). We found 
that when the quality representation switches from $360P$ to $480P$ there
are high bit rate bursts. These bursts cause the erroneous
classification of these segments as $720P$.  In this work, we only trained
the classifier based on the fixed quality switch mode. In future
work, we will consider quality representation switches in our
training.

\subsection{Evaluation of Robustness to Delays and Packet Losses}
\label{Evaluation of robustness to delays and packet losses}

\begin{figure*}[htbp]  
  \subfigure[Streams with different network delays.]{
    \label{fig:safari_delays}
    \includegraphics[height=5cm, width=0.32\textwidth]{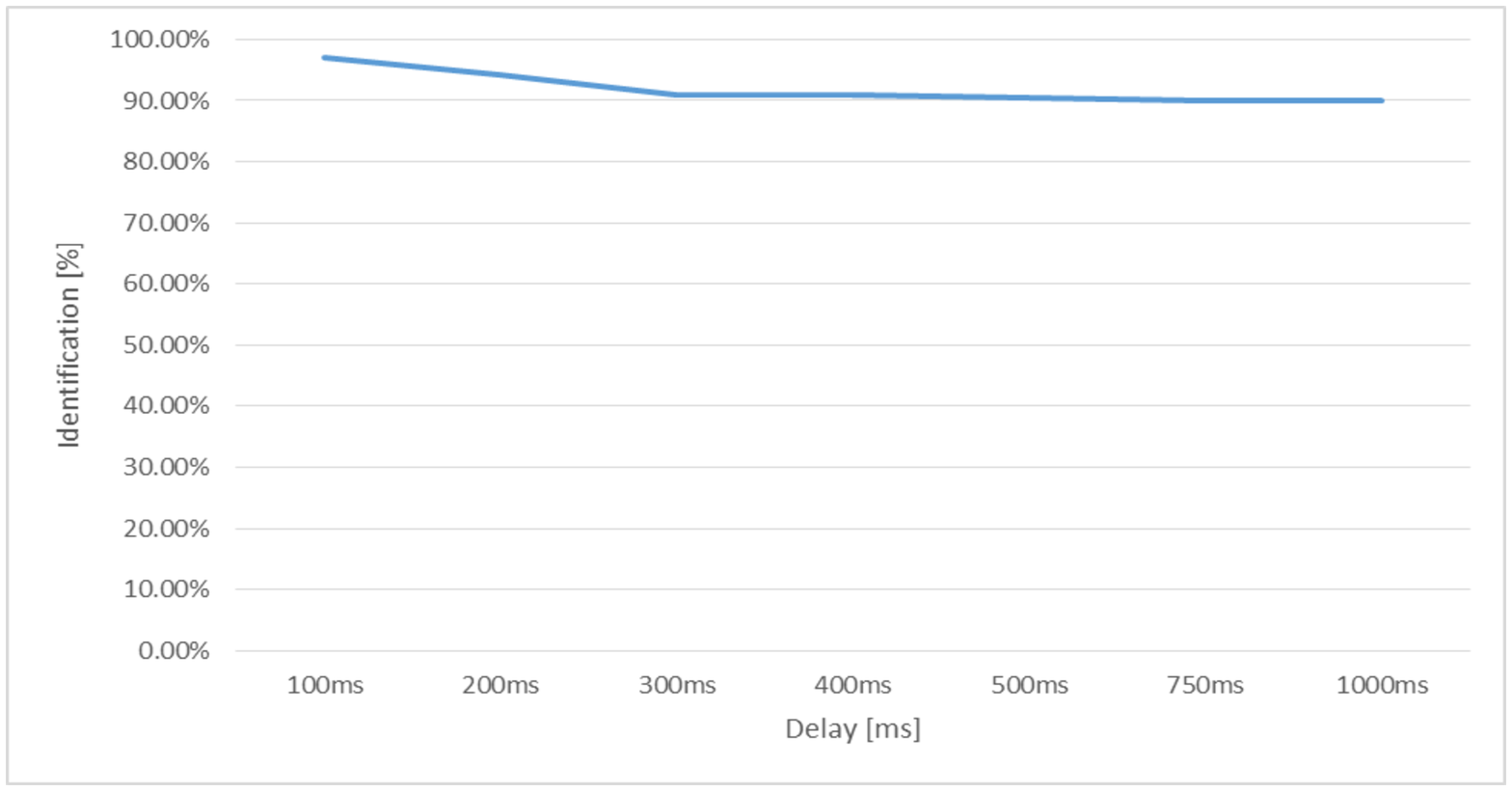}}
  \subfigure[Streams with different percentages of packet loss events.]{
    \label{fig:safari_packet_loss}
    \includegraphics[height=5cm, width=0.32\textwidth]{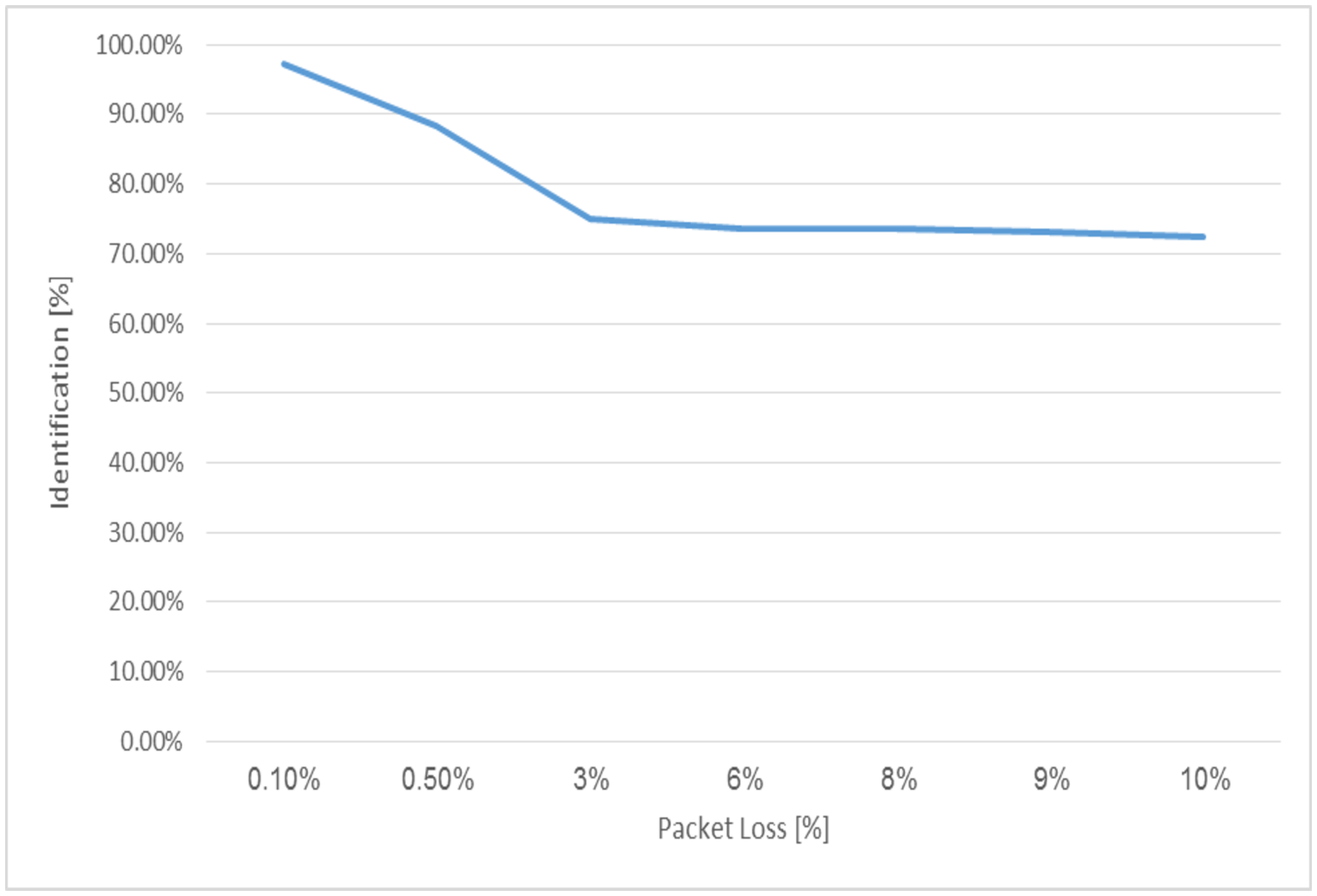}}
  \subfigure[Streams with different combinations of network delays and percentages of packet loss events.]{
    \label{fig:safari_packet_loss_and_delays}
    \includegraphics[height=5cm, width=0.32\textwidth]{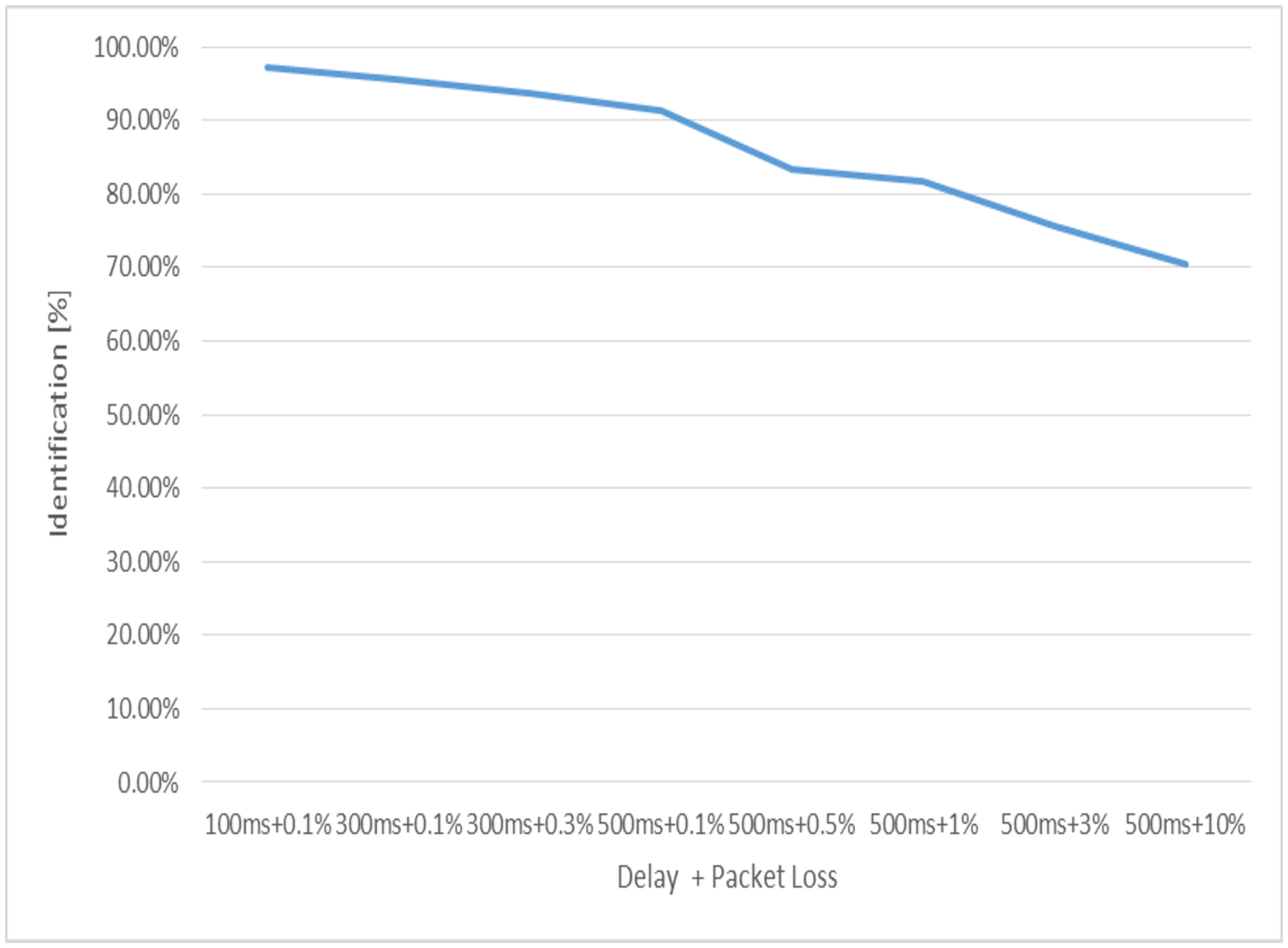}}
\caption{Identification percentage under different network conditions.}
\end{figure*}

Fig. \ref{fig:safari_delays} depicts our algorithm's robustness to
network delays. There was a strong decrease in the classification
accuracy up to $300$ milliseconds delays. Afterward there was a
moderate decrease. The video application QoE is very sensitive to
network delays and delays of over $300$ milliseconds are easily
detected. The overall classification accuracy decreased after $1000$
milliseconds by only $7 \%$.

Fig. \ref{fig:safari_packet_loss} plots our algorithm's robustness to
packet losses. Packet losses of $3 \%$ decreased our classification
accuracy by $20 \%$. We found out that the traffic behavior during
packet loss events was different from our normal testing model.  After
$10 \%$ packet losses (the video is practically halted) our
classification accuracy decreased to $73 \%$.

Fig \ref{fig:safari_packet_loss_and_delays} plots our algorithm's
robustness to combinations of network delays and packet losses.
$500 ms$ delays plus $10 \%$ packet losses decreased our
classification to $70 \%$. However, in real life scenarios it would be
impossible to watch this stream (very low QoE). 

To conclude, our solution (like the other solutions) is somewhat
sensitive to packet losses. Increasing its robustness is left
as future work.

\subsection{User Buffer Estimate}
\label{User buffer estimation}
\begin{figure}[htbp]
\centering{\includegraphics[width= 8cm, height=8cm]{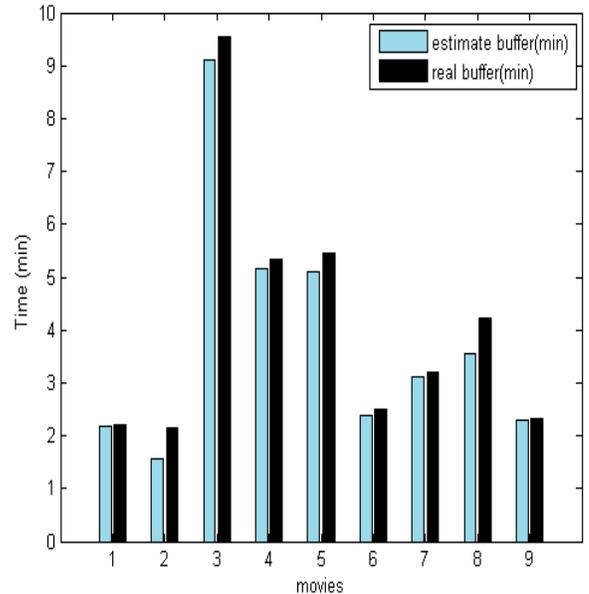}}
\caption{Buffer estimate vs. video duration}
\label{fig:Safari_buf_size}
\end{figure}

Fig. \ref{fig:Safari_buf_size} shows our buffer estimate compared to
the real buffer measurement. The experiments were conducted using the
entire dataset (fixed and auto modes). For simplicity, we present
the total sum. The average estimate drift between the full video
duration and our estimate was $0.276$ seconds and the STD was
$0.25$. The average estimate drift per feature was $0.035$ seconds
with a STD of $0.047$.

\subsection{Classifier Comparisons}
\label{Classifiers comparison}

Our proposed solution is the first classifier for encrypted adaptive
video streaming over HTTPS. In this section, we describe and compare
to two other new classification approaches: a \naive{}
bit rate classifier and an algorithm based on a network
traffic malware fingerprinting algorithm\cite{shimonimalicious}. Since
the malware fingerprinting is not designed for auto representation
switching we used the fixed mode dataset in the tests. The
\naive{} algorithm uses the average bit rate that was calculated from
Fig. \ref{fig:safari_confidence} for each quality. We used our entire
fixed representation testing dataset and found the
closest average quality bit rate for each feature.
Fig. \ref{fig:safari_conf_without_algo} illustrates the \naive{}
approach and the proposed algorithm is presented in
Figs. \ref{fig:safari_confusion_matrix_our_fixed},
\ref{fig:safari_conf_auto_from the data set},
\ref{fig:safari_confusion_matrix_auto_not_from_dataset}. Table \ref{tab:fixedMethodsComparison}
summarize this comparison. It shows that our proposed solution
(based on bit rate) achieved the highest identification results whereas
all the other algorithms using time differences obtained much lower
identification results.


\begin{table}[htbp]
  \centering
  \begin{tabular}{| l | l | l|}\hline	
	Feature & classifier & average confusion \\ \hline
	Bit rate & Naive bit rate & $88.23 \%$\\\hline
	Time differences & Shimoni et al. \cite{shimonimalicious} & $38.26 \%$\\\hline
	Bit rate & Shimoni et al. \cite{shimonimalicious} & $81.46 \%$\\\hline
	Time differences & Proposed solution & $62.21 \%$\\\hline
	Bit rate & Proposed solution & $97.18 \%$\\\hline
  \end{tabular}
  \caption{Comparison of the different classifiers and feature
    creation methods on the \testFixed{} dataset. Note that the \naive{} algorithm is based on bit rate features and cannot be used with time differences.}
  \label{tab:fixedMethodsComparison}
\end{table}

\section{Conclusions}
\label{Conclusions}
We propose a novel algorithm for YouTube HTTP adaptive video
streaming quality representation classification. Our solution was
tested on the Safari (Flash player) browser with offline and online
network traffic over HTTPS. We achieved an average classification
accuracy of $97.18 \%$ in the fixed mode and $97.14 \%$ in the automatic
quality representation switching mode. The algorithm estimates the user
buffer playout level after each segment download with an average error
of $0.035$ seconds. The proposed solution exhibited $8.95 \%$ better
average classification results than a \naive{} classifier
approach. In this work we used the one-dimensional bit rate feature.
We showed that our solution is more vulnerable to packet losses than
to network delays. Adding features to strengthen robustness to
packet losses is one of our future goals.
	
The DASH encrypted traffic quality representation classification
problem still faces many challenges. In this work, we presented
YouTube with the Safari browser over HTTPS as a use
case. Classification of other browsers' streaming is one of our future
goals. The Chrome and Safari auto modes have similar network traffic
behavior but our experiments suggest that the Safari dataset is not similar
enough (the same videos have different total bit rates) to achieve
high accuracy results; thus new datasets for Chrome (fixed/auto) are
needed. The use of state-of-the-art network transport protocols such
as HTTP$2$/SPDY and QUIC that have multiplexed connections should be
investigated. TOR traffic morphing may also be a challenge to
statistical classification \cite{wright2009traffic}. However, we cannot
confirm that this is a problem since in our testing the videos failed
to play smoothly even in $360P$.


\bibliographystyle{unsrt}
\bibliography{report}   

\begin{thebibliography}{10}

\bibitem{Cisco_2}
Cisco.
\newblock Cisco visual networking index: Global mobile data traffic forecast
  update, 2012-2016, 2012.

\bibitem{Cisco_zettabytes}
Cisco.
\newblock The zettabyte era: Trends and analysis, 2015.

\bibitem{sandvine_2014}
Sandvine.
\newblock Sandvine global internet phenomena report h1 , 2014, 2014.

\bibitem{GoogleSSL}
Google.
\newblock Google webmaster central blog: Https as a ranking signal, august,
  2014, 2014.

\bibitem{AdCreator}
Celtra Inc.
\newblock {AdCreator Now Brings Video Ad Content Into Focus}, April 2013.

\bibitem{dainotti2012issues}
A.~Dainotti, A.~Pescape, and KC. Claffy.
\newblock Issues and future directions in traffic classification.
\newblock {\em Network, IEEE}, 26(1):35--40, 2012.

\bibitem{valenti2013reviewing}
S.~Valenti, D.~Rossi, A.~Dainotti, A.~Pescap{\`e}, A.~Finamore, and M.~Mellia.
\newblock Reviewing traffic classification.
\newblock In {\em Data Traffic Monitoring and Analysis}, pages 123--147.
  Springer, 2013.

\bibitem{cao2014survey}
Z.~Cao, G.~Xiong, Y.~Zhao, Z.~Li, and L.~Guo.
\newblock A survey on encrypted traffic classification.
\newblock In {\em Applications and Techniques in Information Security}, pages
  73--81. Springer, 2014.

\bibitem{dubin2012progressive}
R.~Dubin, O.~Hadar, A.~Noam, and R.~Ohayon.
\newblock Progressive download video rate traffic shaping using tcp window and
  deep packet inspection.
\newblock In {\em WORLDCOMP}, May 2012.

\bibitem{niemczyk2014identification}
B.~Niemczyk and P.Rao.
\newblock Identification over encrypted channels.
\newblock In {\em BlackHat USA}, Aug. 2014.

\bibitem{ssl_clas}
P.~Fu, L.~Guo, G.~Xiong, and J.~Meng.
\newblock Classification research on ssl encrypted application.
\newblock In {\em Trustworthy Computing and Services}, volume 320 of {\em
  Communications in Computer and Information Science}, pages 404--411. Springer
  Berlin Heidelberg, 2013.

\bibitem{SSL_ext}
P.~Fu, G.~Xiong, Y.~Zhao, M.~Song, and P.~Zhang.
\newblock An identification method based on ssl extension.
\newblock In {\em Symposium on Research in Attacks, Intrusions and Defenses},
  pages 1--6, 2013.

\bibitem{korczynski2012classifying}
M.~Korczynski and A.~Duda.
\newblock Classifying service flows in the encrypted skype traffic.
\newblock In {\em IEEE International Conference on Communications (ICC)}, pages
  1064--1068. IEEE, June 2012.

\bibitem{DASH_RFC_1}
ISO/IEC.
\newblock {Information technology - Dynamic adaptive streaming over HTTP
  (DASH)}, May 2014.

\bibitem{paxson1994empirically}
Vern Paxson.
\newblock Empirically derived analytic models of wide-area tcp connections.
\newblock {\em IEEE/ACM Transactions on Networking (TON)}, 2(4):316--336, 1994.

\bibitem{alshammari2010unveiling}
R.~Alshammari and AN. Zincir-Heywood.
\newblock Unveiling skype encrypted tunnels using gp.
\newblock In {\em IEEE Congress on Evolutionary Computation (CEC)}, pages 1--8.
  IEEE, July 2010.

\bibitem{zander2005self}
S.~Zander, T.~Nguyen, and G.~Armitage.
\newblock Self-learning ip traffic classification based on statistical flow
  characteristics.
\newblock In {\em Passive and Active Network Measurement}, pages 325--328.
  Springer, 2005.

\bibitem{zhang2010identification}
D.~Zhang, C.~Zheng, H.~Zhang, and H.~Yu.
\newblock Identification and analysis of skype peer-to-peer traffic.
\newblock In {\em Fifth International Conference on Internet and Web
  Applications and Services (ICIW)}, pages 200--206, May 2010.

\bibitem{paredes2012practical}
I.~Paredes-Oliva, I.~Castell-Uroz, P.~Barlet-Ros, X.~Dimitropoulos, and
  J.~Sole-Pareta.
\newblock Practical anomaly detection based on classifying frequent traffic
  patterns.
\newblock In {\em IEEE Conference on Computer Communications Workshops (INFOCOM
  WKSHPS)}, pages 49--54, March 2012.

\bibitem{bonfiglio2009detailed}
D.~Bonfiglio, M.~Mellia, M.~Meo, and D.~Rossi.
\newblock Detailed analysis of skype traffic.
\newblock {\em Multimedia, IEEE Transactions on}, 11(1):117--127, 2009.

\bibitem{chen2006quantifying}
KT. Chen, CY. Huang, P.~Huang, and CL. Lei.
\newblock Quantifying skype user satisfaction.
\newblock In {\em ACM SIGCOMM Computer Communication Review}, pages 399--410.
  ACM, 2006.

\bibitem{hjelmvik2009statistical}
E.~Hjelmvik and W.~John.
\newblock Statistical protocol identification with spid: Preliminary results.
\newblock In {\em Swedish National Computer Networking Workshop}, May 2009.

\bibitem{bar2010realtime}
R.~Bar-Yanai, M.~Langberg, D.~Peleg, and L.~Roditty.
\newblock Realtime classification for encrypted traffic.
\newblock In {\em Experimental Algorithms}, pages 373--385. Springer, May 2010.

\bibitem{white2011phonotactic}
AM. White, AR. Matthews, KZ. Snow, and F.~Monrose.
\newblock Phonotactic reconstruction of encrypted voip conversations: Hookt on
  fon-iks.
\newblock In {\em IEEE Symposium on Security and Privacy (SP)}, pages 3--18.
  IEEE, May 2011.

\bibitem{wright2007language}
CV. Wright, L.~Ballard, F.~Monrose, and GM. Masson.
\newblock Language identification of encrypted voip traffic: Alejandra y
  roberto or alice and bob?
\newblock In {\em USENIX Security}, page~3, 2007.

\bibitem{SiboniCohen2014}
S.~Siboni and A.~Cohen.
\newblock Botnet identification via universal anomaly detection.
\newblock In {\em IEEE Workshop on Information Forensics and Security (WIFS)},
  pages 101--116. IEEE, Dec. 2014.

\bibitem{shimonimalicious}
A.~Shimoni and S.~Barhom.
\newblock Malicious traffic detection using traffic fingerprint.
\newblock \url{https://github.com/arnons1/trafficfingerprint}, 2014.

\bibitem{Lempel1978}
Avraham Lempel and Jacob Ziv.
\newblock Compression of individual sequences via variable-rate coding.
\newblock {\em IEEE Transmissions on Information Theory}, 24(5):530 -- 536,
  September 1978.

\bibitem{lawrencefiddler}
{Fiddler-The Free Web Debugging Proxy by Telerik}.
\newblock \url{http://www.telerik.com/fiddler}, 2012.

\bibitem{ourDB}
{Dataset of the Paper}.
\newblock \verb#https://drive.google.com/foldervie# \newline
  \verb#w?id=0B_NMAPuEyaa6flNRcUY2QnVVWU1FczdZWEJRbDMzT0# \newline
  \verb#9zSkd6T3FReHhRVndmNmVyaDcyQjA&usp=sharing#.

\bibitem{arthur2007k}
D.~Arthur and S.~Vassilvitskii.
\newblock k-means++: The advantages of careful seeding.
\newblock In {\em Proceedings of the eighteenth annual ACM-SIAM symposium on
  Discrete algorithms}, pages 1027--1035, Jan. 2007.

\bibitem{wright2009traffic}
CV. Wright, SE. Coull, and F.~Monrose.
\newblock Traffic morphing: An efficient defense against statistical traffic
  analysis.
\newblock In {\em NDSS}, 2009.

\end{thebibliography}

\end{document}